\documentclass[twocolumn,nofootinbib]{revtex4-1}

\pdfoutput=1

\usepackage{slashed}

\usepackage[normalem]{ulem}

\usepackage{color}

\usepackage{epsfig}
\usepackage{epstopdf}
\usepackage{epsf}
\input{epsf.sty}
\usepackage{graphicx,amsmath,amssymb}
\usepackage{epsfig}
\allowdisplaybreaks

\def\ei{\end{itemize}}
\def\be{\begin{equation}}
\def\ee{\end{equation}}
\newcommand{\bea}{\begin{eqnarray}}
\newcommand{\eea}{\end{eqnarray}}

\usepackage{bbm,bm,amsmath,amssymb}

\usepackage{hyperref}

\newcommand{\cG}{\mathcal{G}}

\newcommand{\cH}{\mathcal{H}}
\newcommand{\cL}{\mathcal{L}}

\newcommand{\cN}{\mathcal{N}}

\newcommand{\cV}{\mathcal{V}}

\newcommand{\N}{\mathcal{N}}

\def\E{{$E_{7(7)}$}}

\newcommand{\rf}[1]{(\ref{#1})}
\usepackage{float}

\begin{document}

\title{Explaining enhanced UV divergence cancellations}
\author{Renata Kallosh}
\email{kallosh@stanford.edu}
\affiliation{Stanford Institute for Theoretical Physics and Department of Physics, Stanford University, Stanford,
CA 94305, USA}

 \begin{abstract}
 We study supergravities with ``enhanced UV divergence cancellations''.  We show that {\it all  these cancellations  are explained by a simple dimensional analysis of nonlinear local supersymmetry} (NLS). We also show that in all cases where E7-type duality was used in the past via vanishing single scalar limit (SSL) to explain/predict  UV cancellations one could have used  dimensional analysis of NLS. The  SSL  constraints in $d=4$ predict  $L\leq \cN-2$ for UV finiteness,  dimensional analysis of  NLS    predicts $L\leq \cN-1$ for UV finiteness, including enhanced cases like $\cN=5, L=4$.   \end{abstract}

\maketitle


\section{Introduction} 

There are two ways to use nonlinear local supersymmetry\footnote{Nonlinear local supersymmetry is a {\it standard local  supersymmetry of the supergravity Lagrangian}, see for example eqs. (8.21)-(8.25) in  \cite{Cremmer:1979up} in $\cN=8$ supergravity. In amplitudes only the linear part of these nonlinear transformations is manifest. We explained the difference in \cite{Kallosh:2023asd}.} (NLS) in supergravities to study their perturbative UV properties. A simple one is just a dimensional analysis based on properties of a superspace geometry. It is valid for loops of order $L$ where $L<L_{cr}$ to be defined later. In $d=4$ this analysis was performed in \cite{Kallosh:1980fi,Howe:1980th} based on a superspace constructed in \cite{Brink:1979nt, Howe:1981gz}. In $d>4$ this dimensional analysis is performed in \cite{Kallosh:2023css} and we will bring up  the relevant results for enhanced cancellation cases here.

The other one which is  valid for loops of order $L$ where $L\geq L_{cr}$ requires an additional information on supergravities where scalars of the theory are in the ${\cG\over \cH}$ coset space. This additional information is about 
the properties of  candidate counterterms and relation between NLS and duality symmetry groups $\cG$ and local $\cH$-symmetry.  The relevant analysis was performed in $d=4$ in  \cite{Kallosh:2023asd} and described shortly in \cite{Kallosh:2023css} for $d>4$.

In a recent review paper  \cite{Bern:2023zkg} there is a discussion of ``puzzling enhanced ultraviolet cancellations, for which no symmetry-based understanding currently exists''. These 3 supergravity cases are 
\be\label{1}
 \cN=5, L=4, d=4  
\ee
\be
 \cN=4, L=3, d=4 
\ee
\be
 \cN=4, L=2, d=5  
\ee
They are described in \cite{Bern:2014sna} ,   \cite{Bern:2012cd} ,  \cite{Bern:2012gh}, respectively.

Other  UV finite cases in $d=4$ supergravity are explained (or predicted to be UV finite)  by duality E7-type  constraints, see for example \cite{Beisert:2010jx,Freedman:2018mrv}. There are  7 cases in $d=4$  
\be
\hskip 1 cm \cN=8, L=3,4,5,6  
\ee
\be
\hskip 0.4 cm \cN=6, L=3, 4  
 \ee
 \be
 \cN=5, L=3 \label{10}\ee
 
\noindent In  \cite{Kallosh:2023asd,Kallosh:2023css} the cases in \rf{1}-\rf{10} including enhanced cancellations present  a relatively simple part of   a more general study of NLS.   
Therefore,  for the benefit of the reader, we review in this note {\it only the cases of dimensional analysis of NLS predictions}, including all enhanced ultraviolet cancellations, separately from  more general cases with $L\geq L_{cr}$  studied in \cite{Kallosh:2023asd} for  $d=4$, and  in \cite{Kallosh:2023css} for $d\geq 4$.

The purpose of this note is to show that the symmetry explaining these cancellation is a nonlinear local supersymmetry, see Secs. \ref{sec:II}, \ref{sec:III}, \ref{sec:IV}. In Sec. \ref{sec:V} we have added 7 cases of prediction of  UV finiteness based on E7 type SSL limits
 \cite{Beisert:2010jx,Freedman:2018mrv}.
And we added 3 more cases $L= \cN-1$ using  NLS,  
\be
\cN=8, L=7
\ee
\be
\cN=6,  L=5 
\ee    
\be\cN=5,  L=4 \ee 
The last one is in the group of enhanced cancellations  in eq. \rf{1}. 
In Sec.  \ref{sec:VI} we discuss all 12 predictions of dimensional  analysis of NLS, including 3 enhanced cases. In Appendix \ref{App:A} we explain why the constraints from a single scalar soft limit are less poweful than the constraints from from E7 symmetry.

\section{\boldmath $L_{cr}(d, \cN)$  in supergravity}\label{sec:II}
We have found in \cite{Kallosh:2023asd,Kallosh:2023css} that these enhanced ultraviolet cancellations are explained by dimensional analysis in NLS.

 The fact that there are linearized candidate counterterms (CT)'s for UV divergences,  as well as nonlinear ones, is known for a very long time \cite{Kallosh:1980fi,Howe:1980th}. The nonlinear ones are geometric, they depend on torsion and curvature components and on superspace covariant derivatives as described  in $d=4$ on shell Lorentz covariant superspace \cite{Brink:1979nt, Howe:1981gz}.
 
 The distinction between these cases is based on dimensions.\footnote{We will use $\cN=8 \, (\cN=4) $ for maximal (half-maximal) supergravities in diverse dimensions.}  In $d=4$ it was known from \cite{Kallosh:1980fi,Howe:1980th} that the nonlinear superinvariants are available at 
 \be \label{cr}
 L\geq L_{cr} = \cN
 \ee
 In general {\it integer dimensions}, where supergravity theories are known, as well as their symmetries \cite{Salam:1989ihk},  the dimensional analysis shows that
 the nonlinear CT's are first available at $L_{cr}$ depending on $d$ and $\cN$   \cite{Kallosh:2023css}.  When ${2\cN\over d-2}$ is an integer, $L_{cr}$ is given by
\be
L_{cr}={2\cN\over d-2}
\label{crInt}\ee
or  (when ${2\cN\over d-2}$ is not an integer)
\be
L_{cr}=\Big [{2\cN\over d-2}\Big ] +1\, .  
\label{cr}\ee
Here $\Big [{2\cN\over d-2}\Big ]$ is the  integer part of a  number, defined as  the part of the number that appears before the decimal. For cases of enhanced cancellations we find, in agreement with \rf{crInt}  
\be
d=4\ , \qquad L_{cr}=\cN\ .
\label{4}\ee
In $d=5, \, \cN=4$ it is, in agreement with \rf{cr} 
 \be
L_{cr}=\Big [{2\cN\over d-2}\Big ] +1\, , \quad \rightarrow \quad   L_{cr}=\Big [{8 \over 3}\Big ] +1 = 3
\label{5}\ee

\section{Dimensional analysis of NLS }\label{sec:III}
Consider first the case of pure gravity.
In \cite{Kallosh:1974yh}  in pure (no matter) gravity the analysis of {\it gauge-independent}   perturbative UV divergences was  performed based on  a dimensional analysis of the {\it on shell} Riemannian space geometry where
\be
R=R_{\mu\nu}=0 \qquad R_{\mu\nu\lambda \delta}\neq 0
\ee
A single 2-loop UV divergence was predicted in \cite{Kallosh:1974yh} based on the fact that in $d=4$ the action
\be
S^{2-loop}= \kappa^2 \int d^4 x  \sqrt {|g|}\, R_{\alpha \beta}{}^{\mu\nu} R_{\mu \nu}{}^{\rho\sigma} R_{\rho\sigma}{}^{\alpha \beta}
\ee
is dimensionless:  -2 (from  $\kappa^2$)  -4 (from $d^{4}x$) +6 (from  $R^3$) =0. The absence of other candidate CT's at 2-loop order in \cite{Kallosh:1974yh} was due to an algebraic  fact that 
only one of all independent scalar invariants  in the four-dimensional Riemannian space can give a non-zero contribution on the mass shell when $R=R_{\mu\nu}=0$.

At higher loops,  candidate CT's   depend on higher powers of the Riemann-Christoffel 4-tensor with insertion of covariant derivatives.  In a symbolic form, the Lagrangian is a scalar, depending on various contraction of Riemann-Christoffel tensors with covariant derivatives. For the $L$-loop the CT is
\be
S^{L}= \kappa^{2(L-1)} \int d^4  x   \sqrt {|g|}\, {\cL} \Big ( ( D_\lambda)^m , (R_{\alpha \beta}{}^{\mu\nu})^n \Big ) ,  
\label{LGR}\ee
and
\be
-2L -2 + m +2n=0\, , \qquad  L={m\over 2} + n -1 \geq 2\nonumber  \ .
\ee
We assumed here that for the S-matrix $n\geq 3$.
At $L=2, m=0, n=3$  we recover the 2-loop case.

Now we look at pure supergravity in the same spirit: the main difference is that we have bosonic and fermionic coordinates, but the on shell geometry is known.
In pure supergravity the dimensional analysis of candidate CT's \cite{Kallosh:1980fi,Howe:1980th} is based on the one shell $d=4$ superspace \cite{Brink:1979nt, Howe:1981gz}. The analog of the $L$-loop candidate CT's 
\rf{LGR} respecting a nonlinear superspace geometry is given by 
\be
CT^{L\geq \cN}=\kappa^{2(L-1)} \int d^4 x \, d^{4\cN}   \theta  \det E \,  \cL (x, \theta)  \ ,
\label{LSGR}\ee
\be
\cL (x, \theta)=   \, \chi_{\alpha \, ijk} \   \chi_{mnl}^\alpha  \, D^{2(L-\cN)}\   \bar \chi_{\dot \alpha} ^{ ijk}  \  \bar \chi^{ \dot \alpha \, mnl}  \ . \label{L}\ee 
Here spinorial superfield $\chi_{\alpha \, ijk}(x, \theta)$ is associated with the superspace torsion, whereas its derivatives define superspace curvatures.
The term $D^{2(L-\cN)} $ in \eqref{L} symbolically denotes multiples of either spinor or spacetime covariant and $\cH$-covariant derivatives with total dimension $2(L-\cN)$.   
These expressions require that the classical equations of motion are valid since the superspace in $\cN\geq 5$ is available only on shell  \cite{Brink:1979nt, Howe:1981gz}. The dimensional analysis requires that
\be
{\rm dim}\,  \cL (x, \theta) = 2 + 2(L-\cN), \   L \geq \cN, 
\ee
 and 
 the superspace Lagrangian $\cL (x, \theta)$ has minimal dimension 2,  the smallest possible dimension for a geometric Lagrangian. A dimension of the volume of superspace is
$
{\rm dim}\,  [d^4 x \, d^{4\cN}\theta ]= -4+2\cN
$. We conclude that in $d=4$  candidate CT's with nonlinear local supersymmetry are available  at
$
L \geq \cN \ ,
$
as was shown in \cite{Kallosh:1980fi,Howe:1980th}.  This is a dimensional analysis of NLS in 
 $d=4$.  An analogous dimensional analysis of NLS is available in other dimensions \cite{Kallosh:2023css}.

\section{3 ``puzzling enhanced'' cancellations}\label{sec:IV}

The most interesting case is
\be
\cN=5, L=4, d=4
\ee 
Investigation of this case revealed a  cancellation of UV divergences in a  sum of 82 diagrams  found in \cite{Bern:2014sna}.  
From our perspective, this   cancellation is explained by the simple dimensional analysis of nonlinear local supersymmetry. Indeed,  according to \cite{Kallosh:1980fi,Howe:1980th}, the nonlinear superinvariants are available for $d = 4$ only if $L\geq L_{cr} =  \cN$, see equation  \rf{1}. Thus they are absent  for $\cN=5, L=4$. 

One may only wonder  why this simple fact was not noticed long ago, and for  the last 9 years this cancellation was considered as unexplained    \cite{Bern:2023zkg}. For a more detailed discussion of this issue see  \cite{Kallosh:2023asd}.

The second case in     \cite{Bern:2012cd} is
\be
\cN=4, L=3, d=4
\ee   
This case has UV divergence cancellation since for $d = 4$ the nonlinear superinvariants are available  only if $L\geq \cN$, see equation  \rf{1}, so they are absent  for $\cN=4, L=3$. 
There are subtleties in this case since there are 1-loop   $U(1)$ anomalies in this theory, see \cite{Carrasco:2013ypa} and at the next loop order, namely, at $L=4$, the theory is UV divergent \cite{Bern:2013uka}. We explained in \cite{Kallosh:2023asd} that these anomalies are also anomalies of  NLS.

The third case in  \cite{Bern:2012gh} is 
\be
\cN=4, L=2, d=5
\ee 
This case has UV divergence cancellation  for $d = 5$, in which case one has $L_{cr} =3$, see \rf{5}. That is why the nonlinear superinvariants are available  only if $L\geq  3$, so they are predicted to  be absent for $L = 2$.
There are also subtleties in this case since  and at the next loop order, namely, at $L=3$ the theory is UV divergent \cite{Bern:2014lha}. Moreover, the corresponding UV divergences break NLS.
More details about this example can be found in \cite{Kallosh:2023css}.

\section{7  cancellations in $d=4$  predicted by  SSL of E7}\label{sec:V}

Duality constraints on CT  in $\cN=8, d=4$ supergravities were derived in  \cite{Beisert:2010jx} using effective string theory and a single scalar soft limit (SSL). The conclusion was that cases with 
\be
 \cN=8, L=3,4,5,6  
\ee
 were predicted to be UV finite. Here we notice that in all these cases 
\be
L<L_{cr}-1 = \cN-1= 7
\ee
This gives a simple explanation of the UV finiteness in $\cN=8, d=4, L=3,4,5,6$ from the standpoint of nonlinear supersymmetry.  But nonlinear supersymmetry adds that also the case 
\be
\cN=8, d=4, L=7
\ee
is predicted to be finite if  NLS  is unbroken.

 In  \cite{Freedman:2018mrv} the conclusion of  \cite{Beisert:2010jx} for $\cN=8$ case was confirmed and in addition, the SSL duality constraints of groups of type E7 were studied in $\cN=6,5$. The conclusion was that cases with 
\be 
 \cN=6, L=3,4  \quad  {\rm  and}   \quad \cN=5, L=3 
 \ee
 are predicted to be UV finite. Here we notice that in all these cases 
\be
L<L_{cr} = \cN-1
\ee
This gives a simple prediction of the UV finiteness in $\cN=6, L=3,4$ and $\cN=5, L=3$ from the standpoint of simple NLS.  But it  also adds 2 cases 
\be
\cN=6,  L=5    \quad  {\rm  and}   \quad    \cN=5, L=4
\ee
are predicted to be finite if simple NLS is unbroken. The second case is actually validated by computations in \cite{Bern:2014sna}, the $\cN=6, L=5$ case is just a prediction.

Thus, we have added to the 3 examples of enhanced cancellation discussed in  \cite{Bern:2023zkg}  7 more examples, which in the past were explained via SSL duality constraints. We added 2 more from dimensional  analysis of NSL.
Now, all 12 of these are shown to follow from dimensional  analysis of NLS.  The most nontrivial one is the case $\cN=5, L=4, d=4$ which duality analysis based on SSL  \cite{Beisert:2010jx,Freedman:2018mrv} is not capable to explain.

This case $\cN=5, L=4, d=4$  is nontrivial since  according to \cite{Bern:2023zkg}
the cancellations  require an interplay between most, if not all, 82 diagrams contributing to a superamplitude. Each of the 82 diagrams is linearly  supersymmetric, but the sum cancels due to a  nonlinear local supersymmetry. 

\section{Discussion of 12 predictions of dimensional  analysis of NLS }\label{sec:VI}
Here we stress that  dimensional  analysis of NLS predictions is based on a {\it local} supersymmetry, there is no need to look into a more complicated analysis of  global $\cG$ symmetry, for example  \E\ in $\cN=8$. And because it is a local symmetry, it is just a standard requirement of a fundamental consistency of the theory.

The analysis of 12 ultraviolet loop cancellations in supergravities predicted by dimensional  analysis of NLS includes 
{\it all 3 known as enhanced ultraviolet cancellations}  and another 9. These are 11 cases in $d=4$,
\bea
&&\cN=8, \, L=3,4,5,6,7 \ , \nonumber \\
&&\cN=6, \, L=3,4,5 \ , \nonumber \\
&&\cN=5, \, L=3,4 \ , \nonumber \\
&&cN=4, \, L=3
 \eea
and there is one in $d=5$,
\be
 \cN=4, \, L=2 \ .\nonumber \\
\ee
All of these have the property that
\be
 d=4: ~ L< L_{cr} = \cN\, ,  \quad d=5: ~ \cN=4\, , ~ L<L_{cr} = 3    \ . \nonumber \\
\ee
We have an interesting  statistics here, since many  of these are already confirmed by loop computations and support the case for   unbroken dimensional analysis of  NLS.

The existence of these cases  points out towards  significance of the difference between linearized candidate CT's  and  nonlinear geometric CT's. The nonlinear ones start at $L_{cr}$ and exists at  $L\geq L_{cr}$. The linearized ones exist at $L< L_{cr}$, but when  NLS is unbroken, they do not appear as UV divergences.

The fact that in all these 12 cases there is the same explanation, via  dimensional analysis of NLS, is satisfactory. We have argued in \cite{Kallosh:2023asd} that in amplitudes only linear supersymmetry is manifest, the nonlinear features of supergravities are only partially explored via single scalar soft limits, see for example \cite{Beisert:2010jx,Freedman:2018mrv}. 
But SSL are less powerful than  the  constraints from the  full E7 type $\cG$-symmetry and local $\cH$-symmetry  which are both linearly realized and independent: shift symmetry SSL analysis is useful but incomplete, it predicts UV cancellation for $L<\N -2$ but  missing the case of $L=\cN-1$, see Appendix.

It is important to stress here that in all cases of maximal supergravities in integer dimensions $d>4$ there are known UV divergences  at $L<L_{cr}$ which break  NLS  \cite{Kallosh:2023css}. So far, $d=4$ looks special for maximal supergravities \cite{Kallosh:2023asd,Kallosh:2023css}. 

The half-maximal $d=4, L=3$ and $d=5, L=2$  cases of enhanced cancellation are easily explained by the fact that $L<L_{cr}$ in both cases. Both of these  enhanced cases   have the following features: after enhanced cancellation at the next loop there is a UV divergence, namely $d=4, L=4$ and $d=5, L=3$ are UV divergent, and moreover, the NLS is broken by these UV divergences. In $d=4$ this is known, see  \cite{Carrasco:2013ypa}, \cite{Bern:2013uka}, in $d=5$ we discuss it in \cite{Kallosh:2023css}, based on the computations in \cite{Bern:2014lha}.

In conclusion, 2 out of 3 cases of enhanced UV cancellations are not  long lasting. Both cases of half-maximal supergravity in $d=4,5$ already at the next loop order have UV divergences breaking nonlinear local supersymmetry. In case of  $\cN=5, L=4, d=4$ we still have to learn what is the fate of the next loop order.

The dimensional analysis of NLS does not  constrain the  candidate CT's for $L\geq L_{cr}$ in the on shell Lorentz covariant superspace  \cite{Brink:1979nt, Howe:1981gz}, as shown in \cite{Kallosh:1980fi,Howe:1980th}. The corresponding CT's at $L\geq L_{cr}$ are available at the level of a simple dimensional analysis.  The next step in the analysis of NLS was performed in \cite{Kallosh:2023asd}, where restriction on candidate CT's for $L\geq L_{cr}$  were studied. The conclusion there was, based on investigation of the  action deformed by candidate CT, required by the preservation of   E7-symmetry. It was found that the deformation breaks NSL. 

Therefore unbroken full nonlinear local supersymmetry and E7-symmetry were found in \cite{Kallosh:2023asd} to protect $\cN\geq 5$ supergravities in $d=4$ from UV divergences even at $L\geq L_{cr}$. The computation of $\cN=5, L=5$ case will be the first to test this prediction.
Hopefully, more new loop computations will be performed and help to understand the theoretical analysis based on symmetries of the theory.

\noindent{\bf {Acknowledgments:}} I am grateful to Z. Bern,  H. Elvang, D. Freedman, A. Linde, H.~Nicolai  and especially to  J. J. Carrasco, R. Roiban  and  Y. Yamada  for stimulating discussions. 
 This work is supported by SITP and by the US National Science Foundation grant PHY-2014215.    
 
\appendix
\section{\boldmath Why SSL predicts UV finiteness for $L\leq  \cN-2 $ whereas dimensional  analysis of NLS predicts it for  $L\leq  \cN-1 $ }\label{App:A}
SSL  was used in \cite{Beisert:2010jx,Freedman:2018mrv} to study the restrictions on UV finiteness in $d=4$. The result was that for $L\leq  \cN-2 $ one can predict  UV finiteness. This was one loop order below the most interesting case of enhanced cancellation in $\cN=5, L=4, d=4$.

Meanwhile here we stated that dimensional analysis  of NLS leads to much stronger results. It  predicts the UV finiteness for $L\leq  \cN-1$, which include the  most interesting case of enhanced cancellation in $\cN=5, L=4, d=4$.

The difference can be explained by the fact that SSL is using only part of  a full E7-type duality symmetry, and also a linear approximation of it. Meanwhile  NLS is using a total group $\cG$. 

The vanishing soft single scalar limit  is a requirement on amplitudes originating from the shift symmetry on scalars, which is part of the duality symmetry. 

Consider for example,  \E\, duality in $\cN=8$ supergravity and compare it with the shift symmetry on scalars.
The classical action \cite{Cremmer:1979up,deWit:1982bul} 
with a local $SU(8)$ symmetry before it is gauge fixed   depends on 133 scalars represented by a  56-bein
\begin{eqnarray}\label{gauge}
{\cal V}=\left(
                                        \begin{array}{cc}
                                          u_{ij} {} ^{IJ}& v_{ijKL} \\
                                           v^{klIJ} &  u^{kl}{}_{KL} \\
                                        \end{array}
                                      \right)\  .                                    \end{eqnarray}
                                      The capital indices $I,J$ refer to \E\,  and small  ones $ij$ refer to $SU(8)$. The 56-bein transforms under a local $SU(8)$ symmetry $U(x)$ and a global \E\,  symmetry    $E $   as follows
\be
{\cal V}(x) \rightarrow U(x) {\cal V}(x) E^{-1} \ .
\label{Vtransf}\ee    
These two symmetries are linearly realized and independent. Here $E \in E_{7(7)}$    is in the fundamental 56-dimensional representation where
\begin{eqnarray}\label{E7}
E=\exp \left(
                                        \begin{array}{cc}
                                          \Lambda_{IJ} {} ^{KL}& \Sigma _{IJPQ} \\
                                          \Sigma ^{MNKL} &  \Lambda ^{MN}{}_{PQ} \\
                                        \end{array}
                                      \right)\ .
                                      \end{eqnarray}
Duality symmetry in \rf{E7} consists of a diagonal transformation $\Lambda _{IJ}{}^{KL}= \delta _{[I}{}^{K} \Lambda _{J]}{}^{L} $ where $\Lambda _{J}{}^{L}$ are the generators of the $SU(8)$ maximal subgroup of \E\ with 63 parameters. The off-diagonal part is  self-dual $\Sigma _{IJPQ}= \pm {1\over 4!} \epsilon_{IJPQ MNKL}   \Sigma ^{MNKL}$ and has 70 real parameters.

In $\cN=8$ supergravity in unitary gauge where there are 70 scalars, the local $SU(8)$ symmetry is gauge fixed in the unitary gauge 
\be
\cV=\cV^{\dagger} \ .
\ee
In the unitary gauge there is no distinction anymore between the capital $I,J$ indices of \E\,  and original local $SU(8)$ indices $i,j$. 
 There are 70 independent scalars in  \begin{eqnarray}\label{cVu}
{\cal V}=\left(
                                        \begin{array}{cc}
                                          u_{ij} {} ^{IJ}& v_{ijKL} \\
                                           v^{klIJ} &  u^{kl}{}_{KL} \\
                                        \end{array}
                                      \right)|_{\cV=\cV^{\dagger} }\, \,    \Rightarrow \left(
                                        \begin{array}{cc}
                                         P^{-1/2} & \, \,  - P^{-1/2}   y\\
                                           - \bar P^{-1/2}  \bar y &   \bar P^{-1/2}  \\
                                        \end{array}
                                      \right)    ,             \nonumber                                                        
                                       \end{eqnarray}
where 
\be
P= 1-y\bar y\, , \quad y_{ij, kl}= \phi_{ijmn}\left ( {\tanh \sqrt{{1\over 8}\bar \phi  \phi}\over \sqrt{\bar \phi \phi}}\right) ^{mn}{}_{kl}\  ,\ee                                       
Here $\phi_{ijkl}$ and $\bar \phi^{ijkl}=\pm {1\over 24} \epsilon^{ijklmnpq} \phi_{mnpq}$ transform in 35-dimensional representation of $SU(8)$. These are 70 physical scalars in the unitary gauge.   

The explicit action of duality symmetry in the unitary gauge was presented in \cite{Kallosh:2008ic}. It is a combination of $E^{-1}$    in \rf{Vtransf} and the field dependent  compensating  $U(x)$ transformation preserving the gauge. The infinitesimal $E_{7(7)}$ transformation takes the form
\be
\delta y\equiv y'-y=\Sigma+ y\bar{\Lambda}-\Lambda \bar y-y\bar{\Sigma}y\ . 
\label{sc} \ee 
In the linear approximation  $y={1\over \sqrt 8} \phi$ 
 \be
\delta \phi_{ijkl}  =\Sigma_{ijkl}+ (\phi \bar{\Lambda}-\Lambda \bar \phi)_{ijkl}\ , 
\ee 
 we  are left with a shift of scalars, $\delta^{shift} \phi_{ijkl}  =\Sigma_{ijkl}$, which allows to remove a constant part of the scalar, as well as an $SU(8)$ rotation $\delta^{SU(8)} \phi_{ijkl}  = (\phi \bar{\Lambda}-\Lambda \bar \phi)_{ijkl}$.
 
But in general the \E\ symmetry depends on 133 parameters $\Sigma, \Lambda$  in the unitary gauge, which acts on scalars as in \rf{sc} and also acts on vectors and spinors, as shown in details in \cite{Kallosh:2008ic}. It is therefore not surprising that the restriction even from simple NLS, which respects the full \E\,,  is stronger than the one from SSL acting only on scalars in the linear approximation.

\bibliography{refs}

\providecommand{\href}[2]{#2}\begingroup\raggedright\begin{thebibliography}{10}

\bibitem{Cremmer:1979up}
E.~Cremmer and B.~Julia, ``{The SO(8) Supergravity}'',
\href{http://dx.doi.org/10.1016/0550-3213(79)90331-6}{{\em Nucl. Phys.} {\bf
  B159} (1979)  141--212}.

\bibitem{Kallosh:2023asd}
R.~Kallosh and Y.~Yamada, ``{Deformation of d=4, N\ensuremath{>} 4
  Supergravities Breaks Nonlinear Local Supersymmetry}'',
  \href{http://arxiv.org/abs/2304.10514}{{\tt arXiv:2304.10514 [hep-th]}}.

\bibitem{Kallosh:1980fi}
R.~E. Kallosh, ``{Counterterms in extended supergravities}'',
\href{http://dx.doi.org/10.1016/0370-2693(81)90964-3}{{\em Phys. Lett.} {\bf
  B99} (1981)  122--127}.

\bibitem{Howe:1980th}
P.~S. Howe and U.~Lindstrom, ``{Higher Order Invariants in Extended
  Supergravity}'',
\href{http://dx.doi.org/10.1016/0550-3213(81)90537-X}{{\em Nucl. Phys.} {\bf
  B181} (1981)  487--501}.

\bibitem{Brink:1979nt}
L.~Brink and P.~S. Howe, ``{The $N=8$ Supergravity in Superspace}'',
\href{http://dx.doi.org/10.1016/0370-2693(79)90464-7}{{\em Phys. Lett.} {\bf
  88B} (1979)  268--272}.

\bibitem{Howe:1981gz}
P.~S. Howe, ``{Supergravity in Superspace}'',
\href{http://dx.doi.org/10.1016/0550-3213(82)90349-2}{{\em Nucl. Phys.} {\bf
  B199} (1982)  309--364}.

\bibitem{Kallosh:2023css}
R.~Kallosh, ``{Is d=4 Maximal Supergravity Special?}'',
  \href{http://arxiv.org/abs/2304.13926}{{\tt arXiv:2304.13926 [hep-th]}}.

\bibitem{Bern:2023zkg}
Z.~Bern, J.~J.~M. Carrasco, M.~Chiodaroli, H.~Johansson, and R.~Roiban,
  ``{Supergravity amplitudes, the double copy and ultraviolet behavior}'',
  \href{http://arxiv.org/abs/2304.07392}{{\tt arXiv:2304.07392 [hep-th]}}.

\bibitem{Bern:2014sna}
Z.~Bern, S.~Davies, and T.~Dennen, ``{Enhanced ultraviolet cancellations in
  $\mathcal N=5$ supergravity at four loops}'',
  \href{http://dx.doi.org/10.1103/PhysRevD.90.105011}{{\em Phys. Rev.} {\bf
  D90} (2014) no.~10, 105011},
\href{http://arxiv.org/abs/1409.3089}{{\tt arXiv:1409.3089 [hep-th]}}.

\bibitem{Bern:2012cd}
Z.~Bern, S.~Davies, T.~Dennen, and Y.-t. Huang, ``{Absence of Three-Loop
  Four-Point Divergences in N=4 Supergravity}'',
  \href{http://dx.doi.org/10.1103/PhysRevLett.108.201301}{{\em Phys. Rev.
  Lett.} {\bf 108} (2012)  201301},
\href{http://arxiv.org/abs/1202.3423}{{\tt arXiv:1202.3423 [hep-th]}}.

\bibitem{Bern:2012gh}
Z.~Bern, S.~Davies, T.~Dennen, and Y.-t. Huang, ``{Ultraviolet Cancellations in
  Half-Maximal Supergravity as a Consequence of the Double-Copy Structure}'',
  \href{http://dx.doi.org/10.1103/PhysRevD.86.105014}{{\em Phys. Rev. D} {\bf
  86} (2012)  105014}, \href{http://arxiv.org/abs/1209.2472}{{\tt
  arXiv:1209.2472 [hep-th]}}.

\bibitem{Beisert:2010jx}
N.~Beisert, H.~Elvang, D.~Z. Freedman, M.~Kiermaier, A.~Morales, and
  S.~Stieberger, ``{$E_{7(7)}$ constraints on counterterms in N=8
  supergravity}'', \href{http://dx.doi.org/10.1016/j.physletb.2010.09.069}{{\em
  Phys. Lett.} {\bf B694} (2011)  265--271},
\href{http://arxiv.org/abs/1009.1643}{{\tt arXiv:1009.1643 [hep-th]}}.

\bibitem{Freedman:2018mrv}
D.~Z. Freedman, R.~Kallosh, and Y.~Yamada, ``{Duality Constraints on
  Counterterms in $\mathcal N=5,\ 6$ Supergravities}'',
  \href{http://dx.doi.org/10.1002/prop.201800054}{{\em Fortsch. Phys.} {\bf
  2018} (2018)  1800054},
\href{http://arxiv.org/abs/1807.06704}{{\tt arXiv:1807.06704 [hep-th]}}.

\bibitem{Salam:1989ihk}
A.~Salam and E.~Sezgin, eds., \href{http://dx.doi.org/10.1142/0277}{{\em
  {Supergravities in Diverse Dimensions}: {Commentary and Reprints (In 2
  Volumes)}}}.
\newblock World Scientific, Singapore, 1989.

\bibitem{Kallosh:1974yh}
R.~E. Kallosh, ``{The Renormalization in Nonabelian Gauge Theories}'',
\href{http://dx.doi.org/10.1016/0550-3213(74)90284-3}{{\em Nucl. Phys.} {\bf
  B78} (1974)  293--312}.

\bibitem{Carrasco:2013ypa}
J.~J.~M. Carrasco, R.~Kallosh, R.~Roiban, and A.~A. Tseytlin, ``{On the U(1)
  duality anomaly and the S-matrix of N=4 supergravity}'',
  \href{http://dx.doi.org/10.1007/JHEP07(2013)029}{{\em JHEP} {\bf 07} (2013)
  029},
\href{http://arxiv.org/abs/1303.6219}{{\tt arXiv:1303.6219 [hep-th]}}.

\bibitem{Bern:2013uka}
Z.~Bern, S.~Davies, T.~Dennen, A.~V. Smirnov, and V.~A. Smirnov, ``{Ultraviolet
  Properties of N=4 Supergravity at Four Loops}'',
  \href{http://dx.doi.org/10.1103/PhysRevLett.111.231302}{{\em Phys. Rev.
  Lett.} {\bf 111} (2013) no.~23, 231302},
\href{http://arxiv.org/abs/1309.2498}{{\tt arXiv:1309.2498 [hep-th]}}.

\bibitem{Bern:2014lha}
Z.~Bern, S.~Davies, and T.~Dennen, ``{The Ultraviolet Critical Dimension of
  Half-Maximal Supergravity at Three Loops}'',
  \href{http://arxiv.org/abs/1412.2441}{{\tt arXiv:1412.2441 [hep-th]}}.

\bibitem{deWit:1982bul}
B.~de~Wit and H.~Nicolai, ``{N=8 Supergravity}'',
\href{http://dx.doi.org/10.1016/0550-3213(82)90120-1}{{\em Nucl. Phys.} {\bf
  B208} (1982)  323}.

\bibitem{Kallosh:2008ic}
R.~Kallosh and M.~Soroush, ``{Explicit Action of E7 on N=8 Supergravity
  Fields}'', \href{http://dx.doi.org/10.1016/j.nuclphysb.2008.04.006}{{\em
  Nucl. Phys. B} {\bf 801} (2008)  25--44},
  \href{http://arxiv.org/abs/0802.4106}{{\tt arXiv:0802.4106 [hep-th]}}.

\end{thebibliography}\endgroup
\bibliographystyle{utphys}
\end{document}